\documentclass[twocolumn,aps,pra,showpacs]{revtex4}
\usepackage{epsfig,amsmath}

\newcommand\nbar{\bar{n}}
\newcommand\Tr{\mathrm{Tr}}

\begin{document}
\title{Deterministic nonclassicality for quantum mechanical oscillators in thermal states}
\author{Petr Marek, Luk\' a\v s Lachman, Luk\' a\v s Slodi\v cka, Radim Filip}
\affiliation{Department of Optics, Palack\' y University,\\
17. listopadu 1192/12,  771~46 Olomouc, \\ Czech Republic}

\begin{abstract}
Quantum nonclassicality is the basic building stone for the vast majority of quantum information applications and methods of its generation are at the forefront of research. One of the obstacles any method needs to clear is the looming presence of decohorence and noise which act against the nonclassicality and often erase it completely. In this paper we show that nonclassical states of a quantum harmonic oscillators initially in thermal equilibrium states can be deterministically created by coupling it to a single two level system. This can be achieved even in the absorption regime in which the two level system is initially in the ground state. The method is resilient to noise and it may actually benefit from it, as witnessed by the systems with higher thermal energy producing more nonclassical states.
\end{abstract}
\maketitle

\section{Introduction}
Classicality and nonclassicality are concepts which originated in quantum optics \cite{Peres, Mandel}. There, certain quantum phenomena such as coherent states \cite{Glauber}, their statistical mixtures, and their interaction with linear optics elements, can be fully explained by the classical coherence theory.
Such quantum states are now denoted as classical and all other states as nonclassical.
The nonclassicality is not just a nomenclature; it has been recognized as an important resource for the modern quantum technology. It has a very close relation to quantum entanglement: entangled states can be created from nonclassical ones simply by linear energy splitting \cite{Kim,Wang,Wolf,Nha,Vogel2014,Adam,Asboth}. It can be also used in communication to ensure secure transfer of information \cite{Qcommunication}, in metrology to enhance precision of measurements beyond their classical limits \cite{Qmeasurement}, and it is a necessary condition for achieving quantum computation \cite{Qcomp}. That is why, in quantum optics, a significant attention has been given to preparation and manipulation of these states \cite{Qoptics, Haroche}.

The nonclassicality is not limited to quantum optics, it can be present in any continuous variables quantum system and it possesses the same inherent value there. One such class of systems is the mechanical oscillator. For the motional mode of a trapped ion \cite{rmp2}, it is currently possible to implement conditional high-quality coherent preparation of nonclassical states  \cite{Home1,Home2}. Nonclassical states of mechanical systems have been recently prepared also in quantum optomechanics \cite{sqmech1}. There, the preparation of more advanced forms of nonclassicality was also suggested either by remote preparation or by coherent transfer \cite{mech1,mech2,mech3,mech4,mech5,mech6} and the coherent transfer itself has been verified for microwave radiation \cite{mech7}. Interfaces based on similar coherent transfer could be in principle used for a flawless transfer of an arbitrary nonclassical quantum state \cite{int1,int2,int3,int4}. One thing these methods have in common is the need for a coherent pump and coherent measurements. One may therefore wonder if that is a fundamental limitation or if it is possible to deterministically generate strong nonclassical features without them. As a first step in this direction we have recently proposed a toy model for thermally driven generation of mechanical squeezed states and proposed its all-optical simulation \cite{kimin}.

In this paper we point out the strong nonclassical properties of the very common energy conserving Jaynes-Cummings interaction. Some of these properties are well known in the field of quantum optics, where a highly nonclassical states of light in a superconducting cavity may be prepared and processed by sequences of atoms passing through it \cite{Haroche,Rempe}. It has been also shown that Jaynes-Cummings interaction can transform classical coherent light into nonclassical \cite{JCCats, Vogelbook}. As a simple illustrative example, consider an excited two level system interacting with oscillator in a vacuum state. For certain value of interaction parameters, one quantum can be completely transferred to the oscillator leaving it in a nonclassical Fock state. However, this requires coherent control of the system, because thermal excitation cannot cause the excited state to be dominantly populated. In the following we shall assume that such thermal control is unavailable and that the systems can be prepared only in \emph{thermal} equilibrium states. We are going to show that even under these constrains, nonclassical states of the oscillator can be \emph{deterministically} prepared. This extension is not trivial because mixtures of quantum states do not, in general, exhibit  the nonclassical properties of thir constituents \cite{NCmixtures}. Furthermore, the produced nonclassicality is not only robust with regards to the initial temperature and mixedness, in some specific scenarios it increases with it. This is interesting for quantum optics, but it is of practical importance for systems which need to be concerned about cooling. The realization that Jaynes-Cummings interaction can be seen as a thermally autonomous nonclassical process (TANP) will be stimulating for the development of quantum optomechanics with solid state two-level systems \cite{SolidOpto}, which can strongly benefit from the ability to deterministically generate nonclassicality without the need for ground state preparation. We provide a detailed proof-of-principle analysis of this effect and for its feasible experimental verification suggest Klyshko's criteria of nonclassicality \cite{Klyshko}.
Finally, we assess the current versions of trapped-ion experiments and reveal that they are sufficient to demonstrate this thermally-driven and nearly autonomous generation of nonclassical mechanical states.

\section{Spontaneous generation of nonclassicality by Jaynes-Cummings model}
In quantum physics, the elementary interaction between single two-level system and oscillator is an unitary and energy conserving process
described by a full Jayness-Cummings Hamiltonian $H=H_0+H_I$,
where $H_0$ is the free evolution Hamiltonian of the joint system
and $H_I$ describes the interaction between its parts
\cite{JC1,JC2,JC3}. The free evolution Hamiltonian
$H_0=\frac{\hbar\omega\sigma_Z}{2}+\hbar\nu a^{\dagger}a$ depends
on transition frequency $\omega=\frac{E_e-E_g}{\hbar}$ of the
two-level system, where $E_g$ and $E_e$ are the energies of the
ground $|g\rangle$ and the excited $|e\rangle$ states,
respectively, and $\nu$ denotes the frequency of the oscillator.
The operator $\sigma_Z$ is inversion operator for the two-level
system and $a^{\dagger}$ and $a$ are the creation and
annihilation operators for the oscillator. We denote
$\Delta=\omega-\nu$ as the detuning parameter. In the
rotating-wave approximation, Jaynes-Cummings interaction
Hamiltonian $H_I=\hbar g (\sigma_+a+a^{\dagger}\sigma_-)$
describes the processes of energy exchange between the oscillator
and the two-level system, where $g$ is interaction strength and
$\sigma_+,\sigma_-$ are two-level raising and lowering operators,
respectively.

The resonant Jaynes-Cummings interaction with $\Delta = 0$ that coherently runs over a time interval $t$ can be represented by unitary operator
\begin{eqnarray}\label{UJC}
    U_{JC} &=& A_{gg}(t) \otimes |g\rangle\langle g| + A_{ee}(t)\otimes |e\rangle\langle e| \nonumber \\
           &+& A_{eg}(t) \otimes |e\rangle\langle g| + A_{ge}(t)\otimes |g\rangle\langle e|,
\end{eqnarray}
where
\begin{eqnarray}\label{A_operators}
    & A_{gg}(t)=\cos\left(gt\sqrt{n}\right) & A_{eg}(t)= a \frac{\sin\left(gt\sqrt{n}\right)}{\sqrt{n}},  \\
    & A_{ee}(t)=\cos\left(gt\sqrt{n+1}\right) & A_{eg}(t)= - a^{\dag} \frac{\sin\left(gt\sqrt{n+1}\right)}{\sqrt{n+1}}, \nonumber
\end{eqnarray}
and $n = a^{\dag} a$ is the number of quanta operator. Over the short interaction time we consider coupling of all the participating systems to their respective thermal baths, and other sources of decoherence, as negligible. Operators (\ref{A_operators}) govern both the discrete exchange of energy between the oscillator and the two level system represented by the annihilation and creation operations, and the continuous nonlinear transformation of the oscillator state performed by the periodic functions of $\sqrt{n}$ and $\sqrt{n+1}$. In past, these terms have been found responsible for the vacuum Rabi oscillations and the collapse-revival phenomena in the population of two-level systems \cite{ScullyBook,HarocheBook}.

\begin{figure}
\centerline{\psfig{width=8cm,angle=0,file=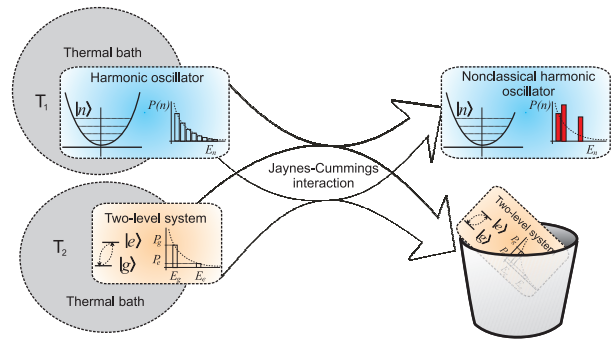}}
\caption{(Color online) The two-level system interacts with the quantum mechanical oscillator and then is discarded. The oscillator is left in a nonclassical quantum state. Both systems were initially in thermal equilibrium states with temperatures $T_1$ and $T_2$.
}\label{scheme1}
\end{figure}

In the following, we shall abandon the two level system and focus on the nonclassicality of the state of the oscillator. Furthermore, we shall assume only a limited control over the entire situation, having both the oscillator and  the two-level system at the beginning in thermal equilibrium states with nonzero temperatures $T_1$ and $T_2$, respectivelly. The situation is schematically depicted in Fig.~\ref{scheme1}. The density operators of the initial thermal states can be expressed as
\begin{equation}\label{rho_TLS}
\rho_{TLS} = p_e|e\rangle\langle e| + (1-p_e) |g\rangle\langle g|
\end{equation}
with $p_e=\frac{\exp(-\frac{\hbar\omega}{k_B T_2})}{1+\exp(-\frac{\hbar\omega}{k_B T_2})}$ for the two level system and
\begin{equation}\label{rho_th}
    \rho_{th} = \sum_{n= 0}^{\infty}\frac{\nbar^n}{(1+\nbar)^{n+1}} |n\rangle\langle n|
\end{equation}
for the mechanical oscillator. The mean excitation $ \nbar $ can be related to the temperature of the system by $\nbar = [ \exp(\frac{\hbar\omega}{k_B T_1}) - 1]^{-1}$.
Under this consideration, the evolution of the oscillator's initial state (\ref{rho_th}) can be obtained by applying a completely positive map as
\begin{equation}\label{}
    \rho_{out} = \sum_{i = g,e}[ p_e A_{ei}(t)\rho_{th} A_{ei}^{\dag}(t) + (1-p_e) A_{gi}(t)\rho_{th}A_{gi}^{\dag}(t)],
\end{equation}
where the respective operators (\ref{A_operators}) can be interpreted as Kraus operators of the transformation. The nature of the operation depends on the initial temperatures of the systems. When the temperatures are equal and $p_e = \nbar/ (1+2\nbar)$, there is no energy exchange and the systems are completely unchanged by the operation. In other cases there is an energy flow between the systems going in the direction of lower temperature. When the two level system is cooler than the oscillator, it absorbs energy from it and vice versa.

The nonclassicality of quantum state is defined as the impossibility to express the state as a mixture of coherent states \cite{Glauber}. In the state produced by the absorption, it can originate from two sources. The first source is the addition of single quanta from the initial excitation of the two level system. This is best seen in the limit case in which the oscillator is initially in the ground state with $\nbar = 0$ and the atom is thermally excited. This results in the oscillator in state
\begin{equation}\label{}
\rho_{out}=p_e\left(\cos^2 gt|0\rangle\langle 0|+\sin^2 gt |1\rangle\langle 1|\right)+(1-p_e)|0\rangle\langle 0|,
\end{equation}
which is nonclassical for any $gt\not=k\pi$ and any $p_e>0$, as can be straightforwardly verified by applying Klyshko's nonclassicality criterion $\frac{P_0P_2}{P_1^2}<\frac{1}{2}$, where $P_k = \langle k|\rho_{out}| k\rangle$, \cite{Klyshko}. If we desire a quantitative statement, we can employ the entanglement potential \cite{Asboth}, which is defined as the amount of entanglement contained in the state
\begin{equation}\label{twomode}
  \rho_{split} =  e^{\frac{\pi}{4}(ab^\dag - a^{\dag} b)} \rho_{out}\otimes|0\rangle\langle 0| e^{-\frac{\pi}{4}(ab^\dag - a^{\dag} b)},
\end{equation}
where $a$ and $b$ are the annihilation operators for the first and the second oscillator mode, respectively. The original optical realization of this operation relies on dividing the optical mode on a balanced beam splitter, but the basic property of the coupling - the inability to create entanglement unless the state is nonclassical - is preserved for any physical system.
The entanglement itself can be quantified with help of logarithmic negativity \cite{Vidal}:
\begin{equation}\label{LN}
   LN(\rho_{split}) = \log_2 \| \rho_{split}^{PT} \|,
\end{equation}
where the superscript $PT$ denotes partial transposition and $\| A \| = \Tr \sqrt{A^{\dag}A}$ represents trace norm. This measure has been originally proposed in \cite{Asboth} and it should be noted that it is not unique. Other measures of nonclassicality \cite{Lee1991, Gehrke2012, Vogel2014} can lead to qualitatively different conclusions. For example, according to \cite{Lee1991}, Fock states are more nonclassical than Gaussian squeezed states, but \cite{Gehrke2012, Vogel2014} suggest otherwise. The entanglement potential itself depends on the particular measure of entanglement employed and the results may differ accordingly. For example, when Gaussian states are considered, logarithmic negativity and entanglement of formation have been shown to be inequivalent measures \cite{Adesso2005}. In the following we shall be using (\ref{LN}) exclusively, because it can be easily numerically evaluated for high-dimensional non-Gaussian states. In order to highlight this choice we shall be calling our entanglement measure \emph{logarithmic negativity potential} (LNP) and using it wherever an amount of nonclassicality is discussed.

In the specific case when the oscillator is in the ground state, we can calculate LNP explicitly as $\mbox{LNP}(\rho_{out})=\log_2 \left[2N(\rho_{out})+1\right]$, where  $\mbox{N}(\rho_{out})=\sqrt{(1-p_e\sin^2 gt)^2+p_e^2\sin^2 gt}-(1-p_e\sin^2 gt)$ \cite{Asboth}. We can see that it indeed is positive for all $gt\not=k\pi$ and any $p_e>0$. Furthermore, if we consider short timescales with $gt \ll 1$, the logarithmic negativity can be expressed as $\mbox{LN}(\rho_{out})\approx \frac{1}{\ln 2}(gt)^4p_e^2$ and we see that it quickly increases with the energy (i.e. the temperature) of the initial two level system. The maximal amount of produced entanglement, which is $\mbox{LN}_{max}=0.5$ corresponding to half of an e-bit, is obtained in the limit of high temperature of the two level system with $p_e = 1/2$. This kind of nonclassicality has been previously discussed and experimentally verified by adding a single quantum to a quantum oscillator in a thermal state \cite{Agarwal1992, Zavatta2007}. Our case is limited because we assume that the two level system can be excited only thermally and as a consequence, perfect addition is impossible and the obtained nonclassicality is lower. On the other hand, in our case there is a second source of nonclassicality which becomes dominant for higher initial temperatures of the oscillator.

This second source of nonclassicality is the modulation of the oscillator's density matrix elements. Over the course of the interaction, the original density matrix with Bose-Einstein thermal statistics (\ref{rho_th}) is transformed into
\begin{eqnarray}\label{rhoout}
    \rho(t) &=& \sum_{n=0}^{\infty} |n\rangle\langle n| \frac{\nbar^n}{(1+\nbar)^{1+n}}   \\
            & & \times \{ p_e[ \cos^2(gt\sqrt{n+1}) + \frac{1+\nbar}{\nbar} \sin^2(gt\sqrt{n})]  \nonumber \\
            & & + (1-p_e)[ \cos^2(gt\sqrt{n}) + \frac{\nbar}{1+\nbar} \sin^2(gt\sqrt{n+1})]\}. \nonumber
            \label{statisticsJC}
\end{eqnarray}
The irrationally oscillating sine and cosine terms reshape the number distribution and can cause some Fock states to be more distinctive, which contributes to the nonclassicality of the state. However, this effect cannot be evaluated as straightforwardly as the previous examples and numerical methods are required for more detailed analysis.
\begin{figure}[t]
\centerline{\includegraphics[width=0.35\textwidth]{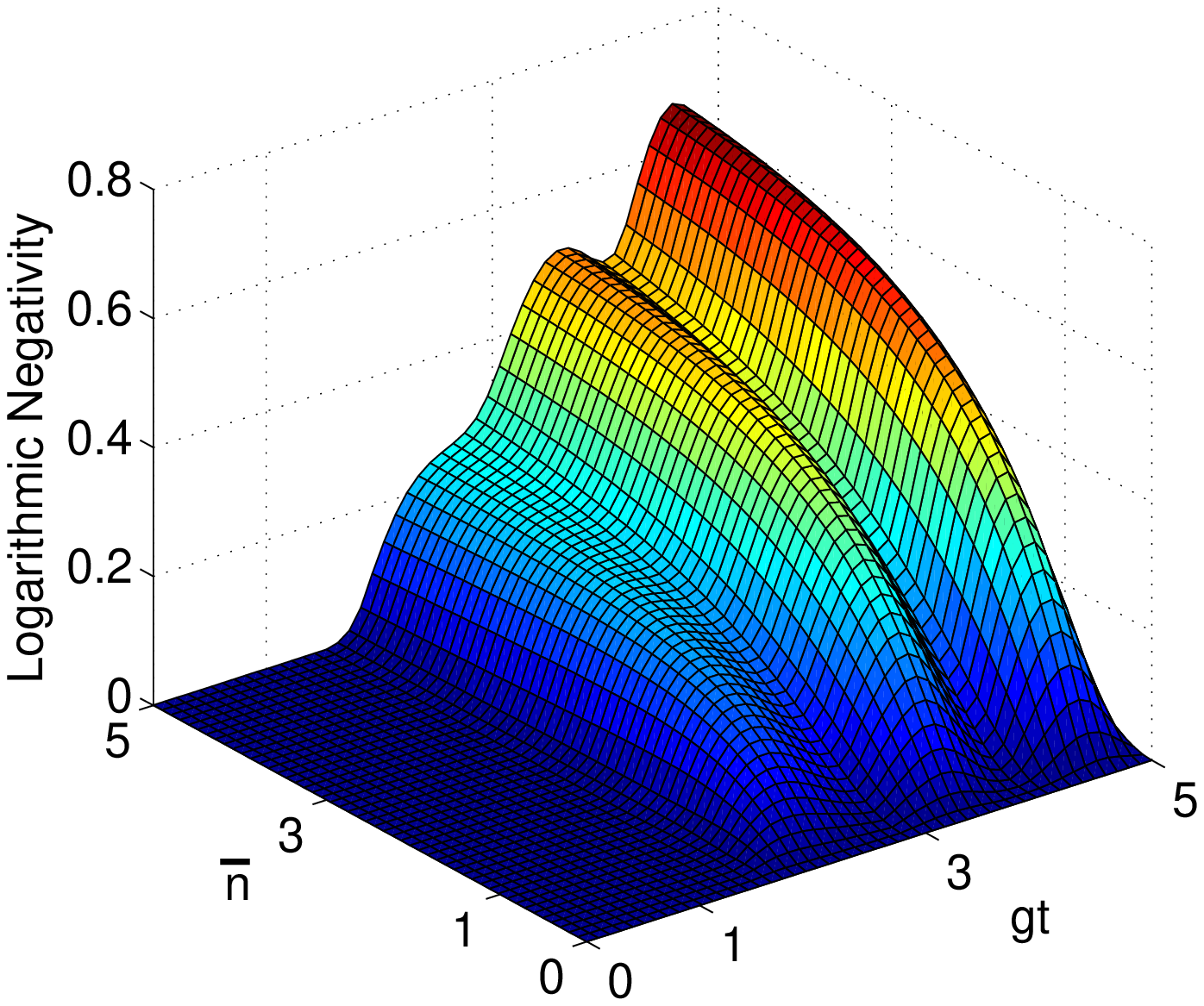}}
\vspace*{-5cm}
\hspace*{-3.5cm}
\textbf{(a)\\}
\vspace*{4.5cm}
\centerline{\includegraphics[width=0.35\textwidth]{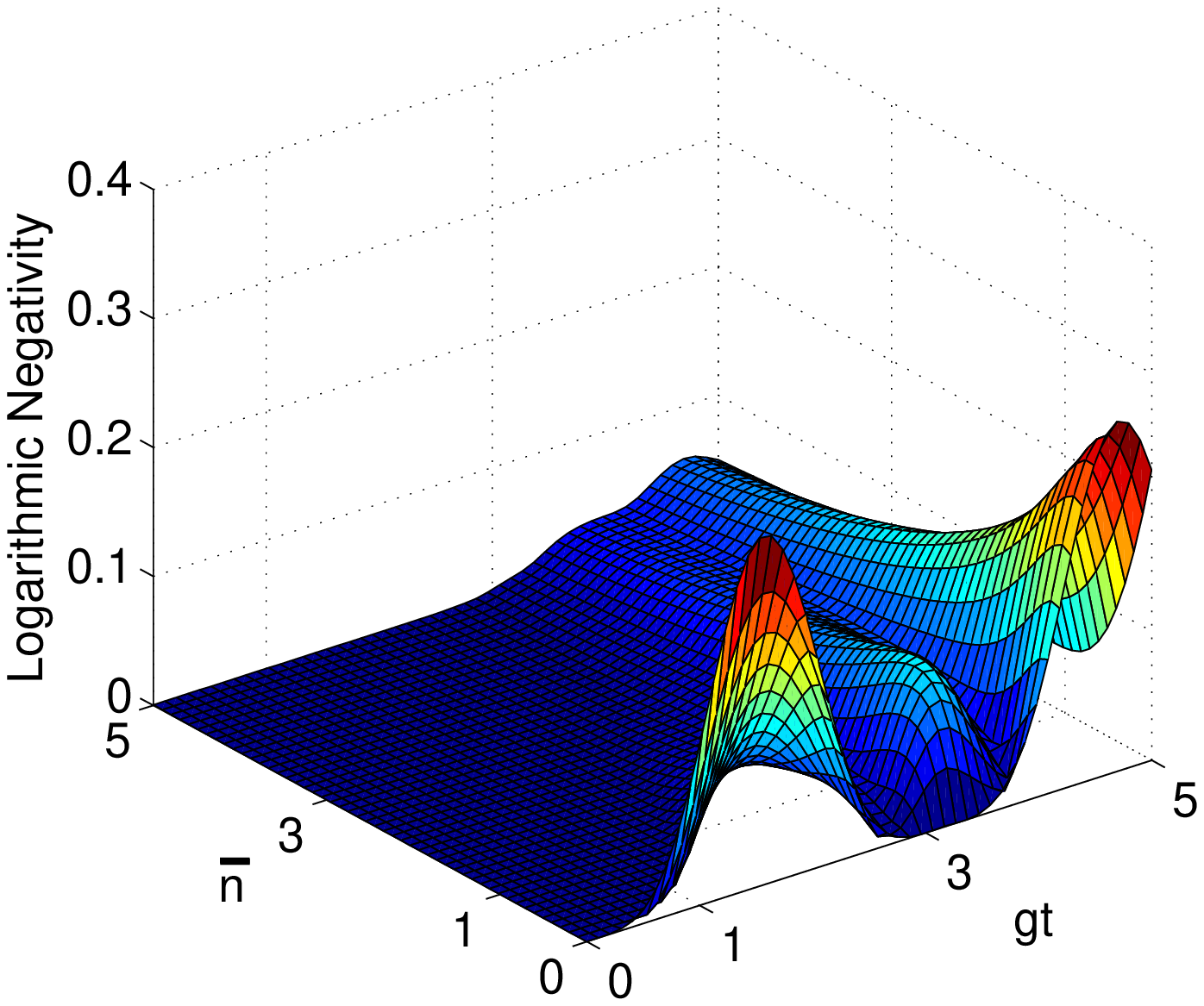}}
\vspace*{-5cm}
\hspace*{-3.5cm}
\textbf{(b)\\}
\vspace*{4.5cm}
\centerline{\includegraphics[width=0.35\textwidth]{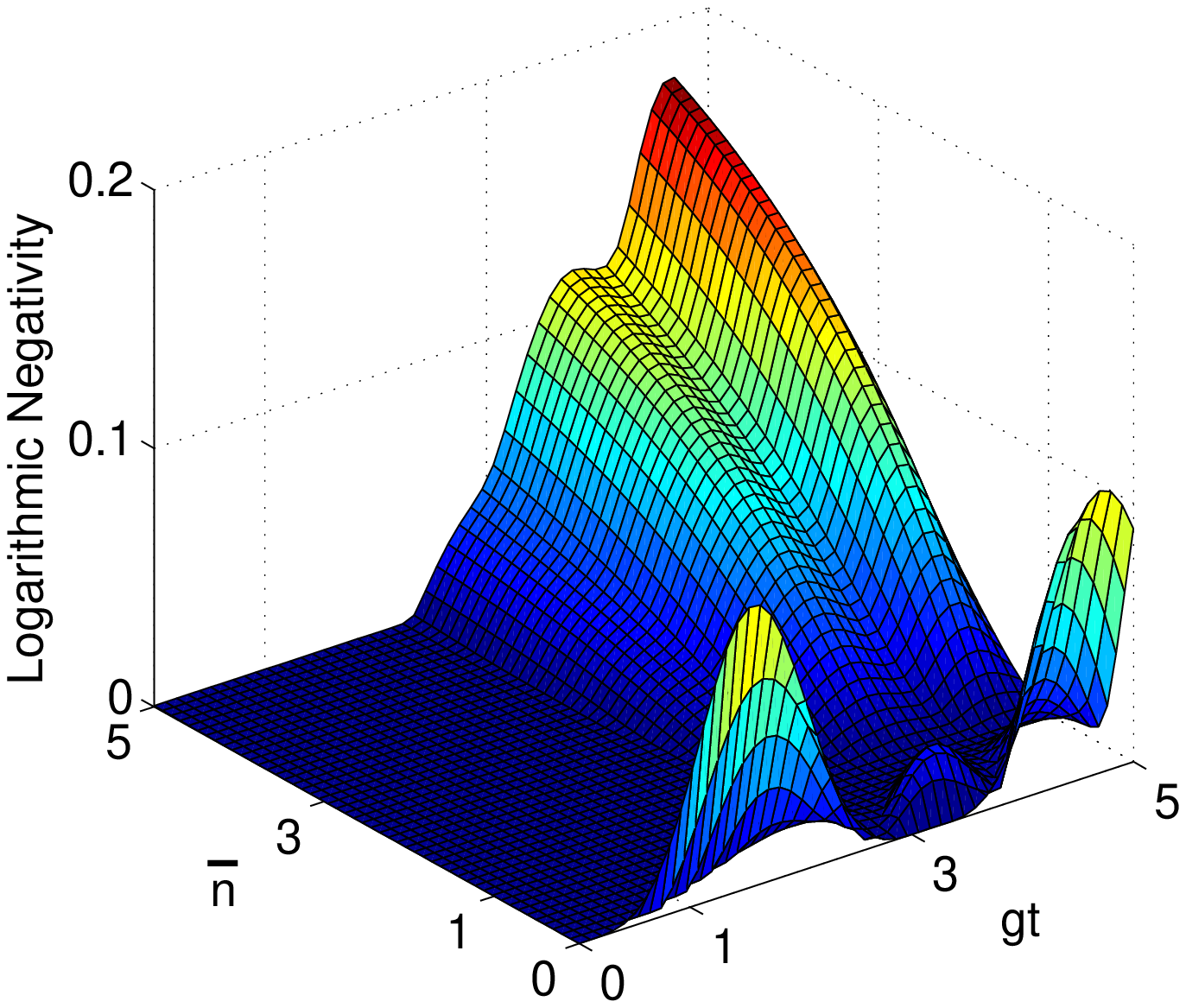}}
\vspace*{-5cm}
\hspace*{-3.5cm}
\textbf{(c)\\}
\vspace*{4.5cm}
\caption{(Color online) The logarithmic negativity potential for a nonclassical state generated from a thermal state with mean excitation of $\nbar$ by Jaynes-Cummings interaction with interaction parameter $gt$. The two level system was initially in ground state (a), maximally thermally excited state with $p_e = 1/2$  (b), and thermally excited state with $p_e = 1/3$.}\label{fig_LN1}
\end{figure}

The amount of LNP present in states generated from thermal states by Jaynes-Cummings interaction with two level systems in various initial states is presented in Fig.~\ref{fig_LN1}. Fig.~\ref{fig_LN1}a shows LNP when the two level system was initially in the ground state and the procedure can be interpreted as a special kind of absorption. Here we can see that the amount of nonclassicality fluctuates with the strength of the interaction, but, more importantly, it grows with the initial energy of the oscillator. This is a very interesting behavior. It demonstrates that the absorption by a single two level system can be interpreted neither as linear absorption also known as loss, nor as subtraction of a single quantum represented by applying annihilation operator $a$, because both of these operations produce no nonclassicality when applied to a thermal state. In our case, the nonclassicality is caused by the oscillating sine and cosine terms and higher initial energy of the thermal state then allows higher Fock states to be affected, stand out, and contribute to nonclassical behavior.

When the two-level system has the maximal thermal energy corresponding to $p_e = 1/2$, which is the situation covered by Fig.~\ref{fig_LN1}b, the NLP is maximal when the oscillator is in its ground state and it decreases with increasing the energy of the oscillator. Finally, Fig.~\ref{fig_LN1}c shows what happens when the two level system is at some intermediate nonzero temperature. We can see that NLP starts positive for the zero energy oscillator and decreases, similarly as in Fig.~\ref{fig_LN1}b. The overall values of NLP are lower, though, which is the consequence of lower proportion of the excited state. The nonclassicality decreases with energy of the oscillator until it reaches a minimal value of zero after which it starts growing again, this time similarly as in Fig.~\ref{fig_LN1}a.

By looking at the trio of enlisted situations side by side, we can see that for obtaining larger nonclassicality we need to couple systems with vastly different energies. This is the reason why there is a local minimum valley in Fig.~\ref{fig_LN1}c and why in Fig.~\ref{fig_LN1}b LNP tends to zero as $\nbar$ increases. Specifically, $p_e = 1/2$ corresponds to infinite temperature, so any increase of $\nbar$ approaches the situation in which the temperature of the two systems is equal.

\begin{figure}[t]
\centerline{\includegraphics[width=0.35\textwidth]{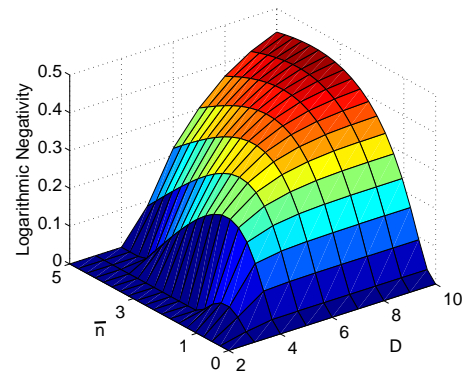}}
\caption{(Color online) The logarithmic negativity relative to mean thermal excitation of the oscillator $\nbar$ and dimensionality of the filter $D$. The interaction constant was chosen as $gt = \pi\sqrt{2}$.  }\label{fig_filterM}
\end{figure}

The generated nonclassicality is arising from abnormally strong presence of certain higher Fock number states. Here, by abnormally strong we mean much higher than could be expected from a mixture of classical states. Higher initial temperature of the oscillator then increases populations of higher Fock states and allows them to be redistributed by the sine and cosine terms of (\ref{rhoout}).This is the reason why the nonclassicality grows with the temperature of the initial thermal state. To illustrate this feature, let us look at how the entanglement of the split state (\ref{twomode}) depends on dimension of the Hilbert space. To do that, we shall consider a projection of the total state (\ref{twomode}) onto a $D^2$ dimensional subspace, which gives us a new density operator
\begin{equation}\label{}
    \rho_D \propto F(D)\otimes F(D) \rho_{split} F(D)\otimes F(D),
\end{equation}
where $F(D) = \sum_{k = 0}^{D} |k\rangle \langle k|$ is a conditional filtering operation representing local scissors.
This probabilistic operation can be used to simulate the effects of imperfect photon number resolving measurements which do not allow us to obtain a full density matrix but only a few of its lowest order elements.
Logarithmic negativity that can be obtained is depicted in Fig.~\ref{fig_filterM} for a single illustrative scenario. We can see that the entanglement can be observed, but as the initial mean energy of the oscillator grows, higher dimension of the density matrix is required. This implicates higher Fock states as those responsible for the generated entanglement and the original nonclassicality.

\begin{figure}[t]
\centerline{\includegraphics[width=0.35\textwidth]{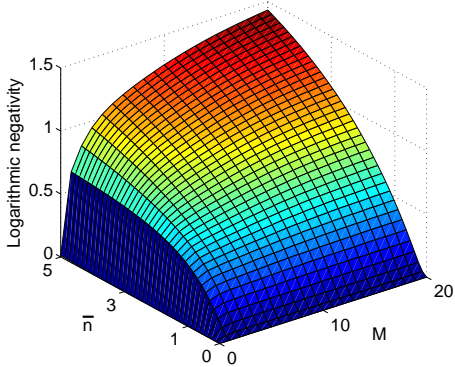}}
\caption{(Color online) The logarithmic negativity relative to mean thermal excitation of the oscillator $\nbar$ and number of times the absorption process is repeated $M$. The interaction constant was chosen as
$gt = \pi$. The absorption process consists of a single interaction with a two level system in the ground state which was subsequently traced over.}\label{fig_Mconverge}
\end{figure}

Another interesting feature, which can be seen in Fig.~\ref{fig_LN1}a, is that for higher thermal energy of the initial state there are two distinct local nonclassicality maxima, which appear when interaction coefficients are $gt = \pi$ and $gt = \pi\sqrt{2}$. These values are not unique and similar trend may be observed whenever the interaction constant is $gt = \pi\sqrt{k}$, where $k$ is an integer. When the two level system is in the ground state and the operation can be therefore interpreted as an absorption, these values ensure that occupancy of some particular number states cannot decrease. As a consequence, when the oscillator is repeatedly exposed to a two-level system in a ground state which absorbs some of its energy and later is traced over, the state of the oscillator
\begin{equation}\label{limitform}
    \rho = \sum_{m = 0}^{\infty} p_m |k m^2\rangle \langle k m^2|
\end{equation}
instead of vacuum. This is a nonclassical state and its nonclassicality expressed by LNP increases with the initial thermal energy and quickly grows with the number of times the absorption is repeated, as can be witnessed by looking at Fig.~\ref{fig_Mconverge}.

\section{Feasible nonclassicality detection}
So far we have been analyzing the nonclassicality of quantum states of oscillators from a sort-of-omniscient point of view. We have been employing the knowledge of full density matrices, which is something that cannot always be fully obtained in realistic conditions. For practical purposes it is therefore necessary to consider nonclassicality criteria, or witnesses, that can be employed with only a partial knowledge of the quantum state. A powerful hierarchy of such sufficient nonclassicality conditions was introduced by Klyshko \cite{Klyshko}. Each of these conditions depends only on a select few diagonal elements of density matrix corresponding to probabilities of finding a specific number of excitations $P_n = \langle n|\rho |n\rangle$. The individual conditions are
\begin{equation}
(n+1) P_{n-1} P_{n+1}-n P_n^2 < 0,
 \label{KHierarchy}
\end{equation}
where $n = 1,2,\ldots$, and none of them can be satisfied by a quantum state that can be expressed as a mixture of coherent states. As a consequence, finding a single satisfied condition decisively marks the scrutinized quantum state as nonclassical.

\begin{figure}[t]
\centerline{\includegraphics[width=0.4\textwidth]{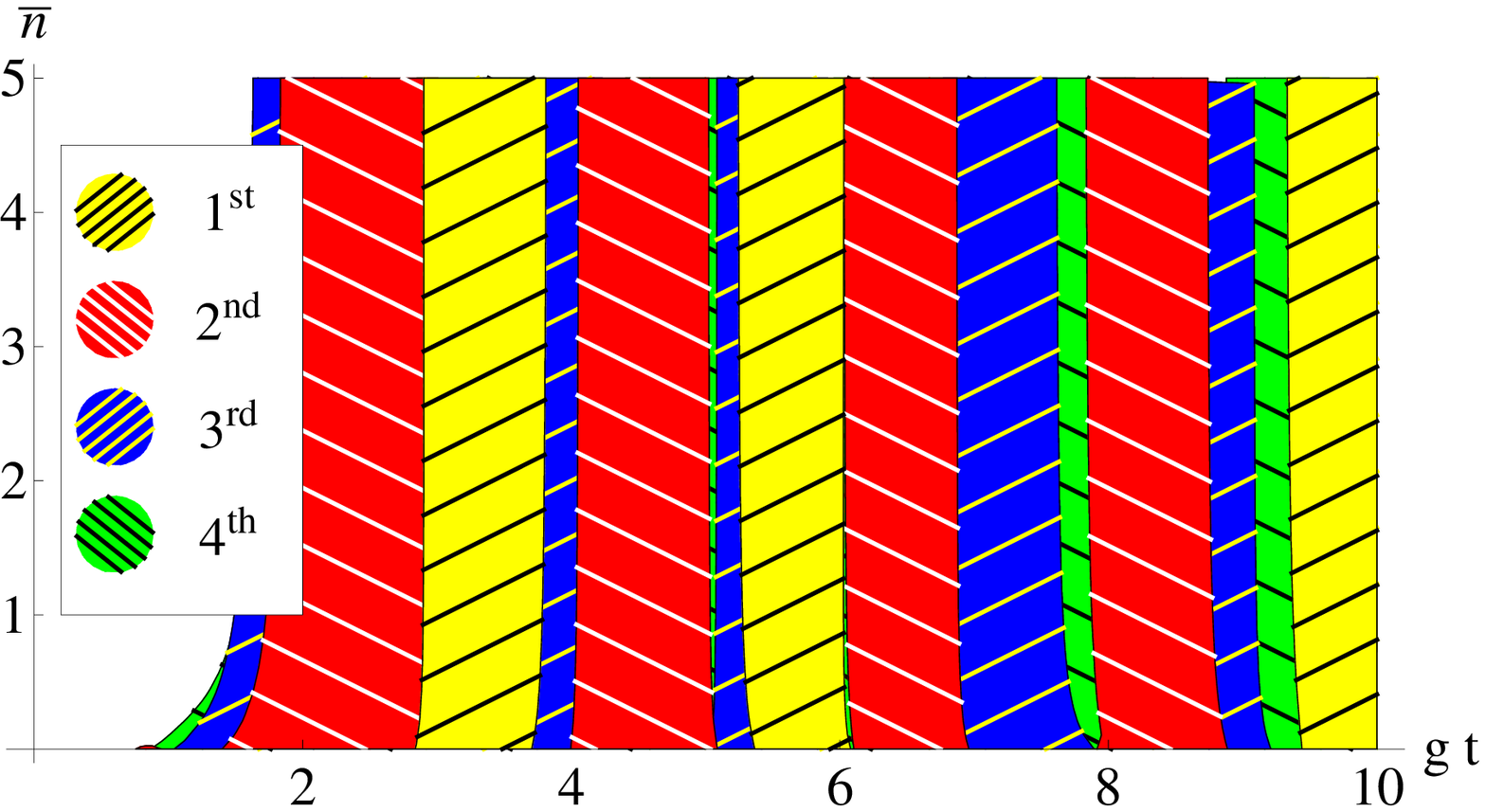}}
\vspace*{-4.5cm}
\hspace*{-5.8cm}
\textbf{(a)\\}
\vspace*{4.5cm}
\centerline{\includegraphics[width=0.4\textwidth]{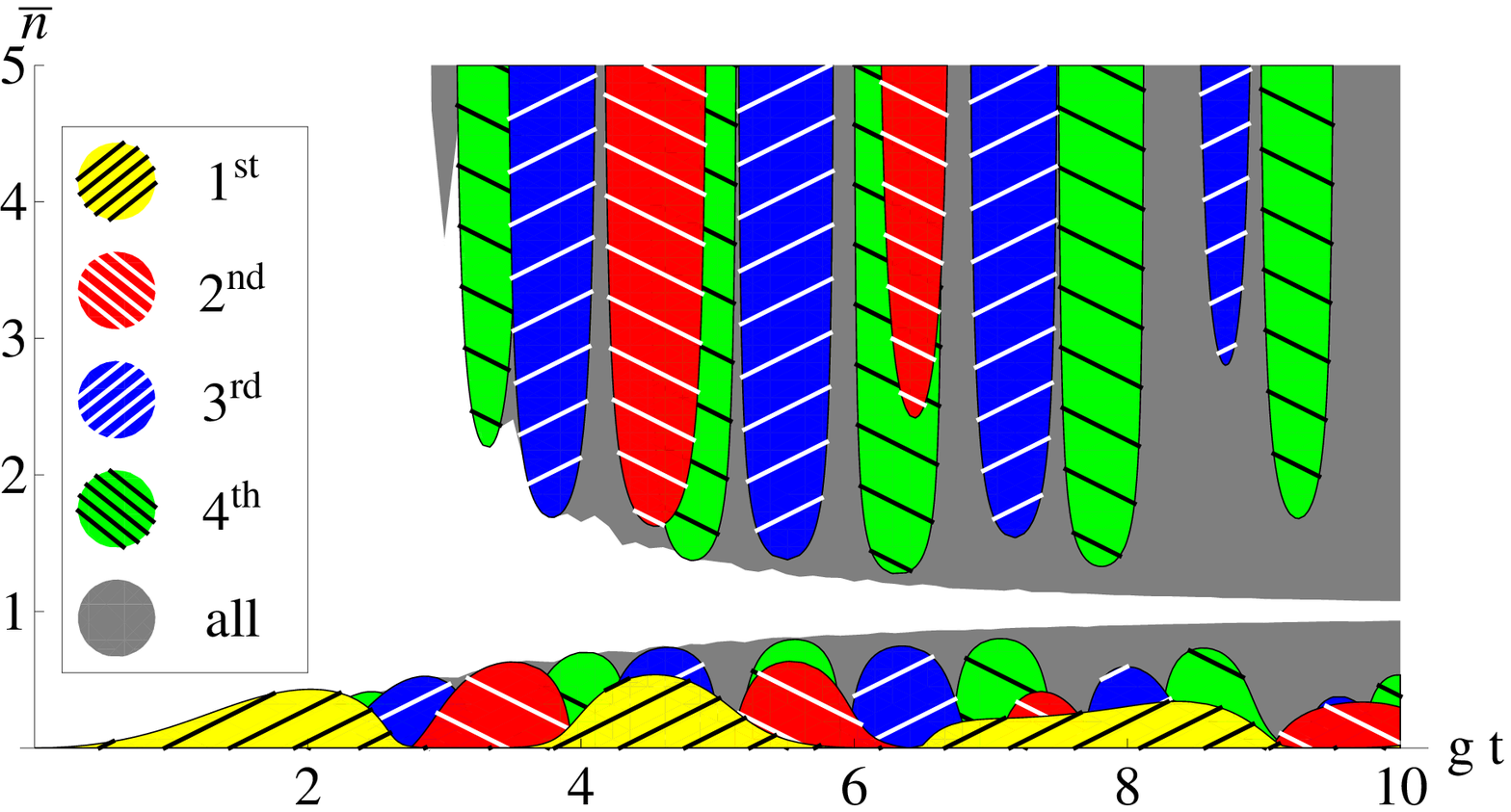}}
\vspace*{-4.5cm}
\hspace*{-5.8cm}
\textbf{(b)\\}
\vspace*{4.5cm}
\caption{(Color online) The nonclassicality of the light after JC as detected by the first four orders in hierarchy (\ref{KHierarchy}). In (a), the two mode system is initially in the ground state, while in (b) the system is in a thermally excited state with $p_e = 1/3$, for which the equivalent energy of the oscillator is $\nbar = 1.$}\label{fig_hierarchy1}
\end{figure}

We can apply this hierarchy of conditions to states produced by the Jaynes-Cummings interaction in order to see which scenarios lead to nonclassicality. Let us start with the perfect saturable absorption where the two level system is initially in its ground state. Fig.~\ref{fig_hierarchy1}a shows that when the total interaction coefficient is larger than some minimal value, roughly $gt \ge \pi/2$, the nonclassicality can be found with just the first three criteria from hierarchy (\ref{KHierarchy}) for arbitrary nonzero values of $\nbar$. This indicates that in a possible experimental realization, the autonomous deterministic generation of nonclassical states from thermal energy can be verified straightforwardly.
Exact numerical tests were performed for values approaching $gt = 50$ and $\nbar = 50$. Outside of this range of parameters we may take advantage of employing approximations. In the limit of high mean excitation $\bar{n} \gg 1$ the Bose-Einstein statistics of the initial state is simplified to $P_n \approx 1/\nbar$.  In this regime, the statistics of excitations of the oscillator (\ref{statisticsJC}) after the interaction is
\begin{equation}
P_n  \approx  \frac{\left(\cos^2 g t \sqrt{n}+\sin^2 g t \sqrt{n+1}\right)}{\bar{n}}=\frac{F_n}{\nbar}
\label{grounApp}
\end{equation}
where function $F_n$ is introduced to abbreviate the notation. Employing hierarchy (\ref{KHierarchy}) gives raise to asymptotic conditions for nonclassicality $n F_n^2-(n+1)F_{n+1}F_{n-1}>0$. This condition can be satisfied for arbitrary $\nbar$, provided we can choose suitable parameters $gt$ and $n$ (see Appendix A for details).

When the two-level system is not in its ground state, the nonclassicality becomes more difficult to detect and it vanishes completely when the temperatures of the system equalize. An example of this can be seen in  Fig.~\ref{fig_hierarchy1}b where, for $p_e = 1/3$, the nonclassicality can be detected in the regions removed from the point of equality ($\nbar = 1$), even though higher orders of the criteria are generally required.

\begin{figure}[t]
\centerline{\includegraphics[width=0.4\textwidth]{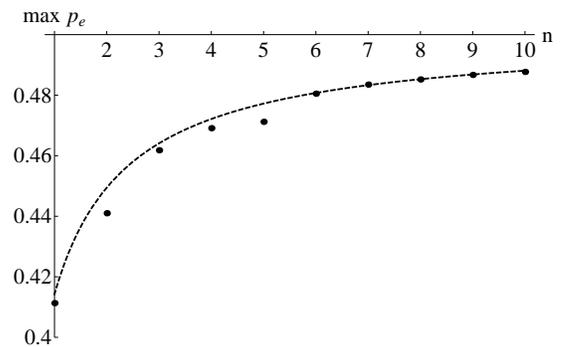}}
\caption{Relation between excitation parameter of the two level system $p_e$ and the order of criteria needed for detection of nonclassicality $n$ in the limit of large initial temperature of the oscillator, $\nbar \rightarrow \infty$. The dashed line corresponds to theoretical prediction (\ref{pCond}). The dots were obtain by numerically searching for maximal $p_e$ that violate criteria of nonclassicality over all possible $gt \in (0,10)$.}\label{fig_hierarchy3}
\end{figure}

The extent to which the absorption by a thermally excited two level system can induce nonclassicality can be again investigated in the limit of high $\nbar$.
Using again the approximation (\ref{grounApp})
the criteria can be expressed as
\begin{eqnarray}
&\ &n\left[2 p_e+F_n(1-2p_e)\right]^2 \nonumber \\
&\ &-(n+1)\left[2 p_e+F_{n-1}(1-2p_e)\right] \nonumber\\
&\times &\left[2 p_e+F_{n+1}(1-2p_e)\right]>0.
\end{eqnarray}
In the regime of maximal violation with $F_n = 2$ and $F_{n-1} = F_{n+1} = 0$ (Appendix A) the criteria then reduce to
\begin{equation}
p_e<\sqrt{n^2+n}-n.
\label{pCond}
\end{equation}
This formula couples the thermal energy of the two level system with the order of criteria needed for detecting some nonclassicality. It was derived under the assumption of large $n$, but numerical analysis reveals that it approximatively holds even for lower orders, see Fig.~\ref{fig_hierarchy3}.

\begin{figure}[t]
\centerline{\includegraphics[width=0.4\textwidth]{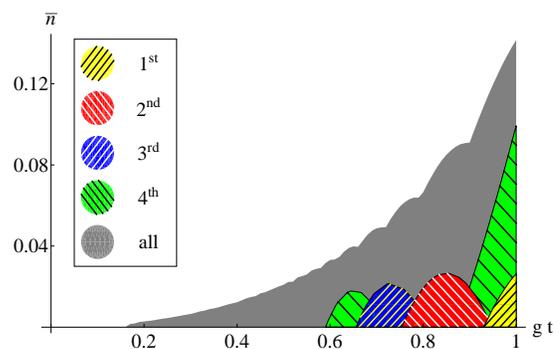}}
\caption{(Color online) The nonclassicality of oscilattor after JC interaction for very small mean energy of the oscillator $\bar{n}$ and coupling strength $g t$. The nonclassicality is detectable even for $g t \ll 1$ but higher order criteria are needed.}\label{fig_hierarchy2}
\end{figure}

Finally, let us look at the opposite regime, where both the interaction strength and the initial thermal energy are small and the two level system is again in the ground state. In this regime, the nonclassicality is difficult to detect and the reason for this is straightforward: when $gt = 0 $ no interaction happens and when $\nbar = 0$ there is no energy in the oscillator to be absorbed and redistributed. However, the nonclassicality can still be detected, it just requires high orders of the hierarchical criteria. In Fig.~\ref{fig_hierarchy2} we can see that while the first three orders of the hierarchy see nonclassicality only in isolated regions, higher orders allow for continuous coverage.
%

\section{Experimental emulation of thermally autonomous nonclassical processes}
The presented calculations suggest that non-classical processes
are likely present in rich variety of thermal physical systems
occurring in nature and might lead to generation of
non-classical states or entanglement in environments with large
thermal energy. Observation of non-classical states of mechanical
oscillators or collective oscillations in mesoscopic solid-state
systems corresponds to a long-standing challenge in a large part
of experimental quantum optics community and proved profitability
of the amount of non-classicality with respect to the initial
oscillator temperature might greatly simplify such pursuits.

\begin{figure}[t]
\centerline{\includegraphics[width=0.4\textwidth]{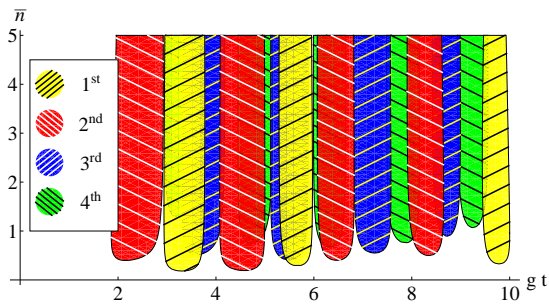}}
\caption{(Color online) The nonclassicality of the single trapped ion's motional state as detected by realistic measurements.
$\bar{n}$ and $gt$ denote initial thermal energy of the ion's motional state and the coupling pulse area representing the strength of the JC interaction, respectively. }\label{simulKlyshko}
\end{figure}

To further stimulate experimental developments based on TANP, it
would be very valuable to emulate the calculated temperature
dependence of TANP in some well controllable and accessible
optomechanical experimental platform. Single ions trapped in Paul
traps seem to be excellent candidates for such task. They offer
great control of the temperature of initial thermal state after
laser cooling stages and recent experimental results show
unprecedent precision of ion's motional state
reconstruction~\cite{Home1,Home2,Hempel2013,Matsukevich}. The
proposed TANP and the corresponding JC interaction can be
straightforwardly implemented by excitation of the ion on a dipole
forbidden transition with a laser frequency detuned from the
resonance by $\omega_T$ to the red, where $\omega_T$ is the
frequency of the ion's motion. If we assume some typical
experimental parameters corresponding to $\omega_T = 2\pi\times
1$MHz, Lamb-Dicke parameter on the employed transition of
$\eta\sim 0.1$ and carrier Rabi frequency $\Omega = 2\pi\times100$kHz   , we
get coupling strength $g \sim \eta \Omega = 2\pi \times$10kHz. The
number of coherent Rabi cycles on first motional sideband for a
single ion can easily reach numbers much higher than considered
$gt=10$ oscillations without any substantial decrease of the
oscillation contrast~\cite{rmp2}, so we can neglect the effect of
finite coherence in the following. The mean phonon number of the
initial thermal state can be well controlled by the effectiveness
and duration of the Doppler and sideband cooling processes with resolution well below single motional quanta
~\cite{slodicka2012, Rossnagel2015}. The non-classicality of the
resulting state after TANP can be conveniently proved by
estimation of the particular phonon
populations~\cite{Meekhof1,Home1} and evaluation of the
corresponding Klyshko non-classicality conditions
(\ref{KHierarchy}). Fig.~\ref{fig_hierarchy1}a suggests that
for excitation pulse areas $gt>2$, the generated non-classicality
should be always detectable for any initial temperature of the
oscillator. However, if we assume common measurement scheme of the
phonon populations $P(n)$ based on driving the first blue sideband
and conservative measurement error on individual population
measurements of the five lowest populations $\sigma_{P(n)} = 0.01$
\cite{Home1,Home2}, the region of the observable non-classicality
slightly reduces, see Fig.~\ref{simulKlyshko}. This is mainly
caused by the high requirements for measurement precision in the
region of the small thermal energy of initial motional state.
Non-classicality regions for higher initial thermal populations
clearly demonstrate how thermal energy entering TANP improves the
observability of the initial generated non-classicality.
In the presented simulation, experimentally feasible regions have been estimated by taking the worst possible case of combinations of individual population probabilities shifted by $\sigma_{P(n)}$ for the given evaluated criterion from hierarchy (\ref{KHierarchy}). We have assumed that experiment operates deep in the Lamb-Dicke regime and so the internal population after the TANP can be reshuffled using some auxiliary electronic level to the ground state of the two-level system without any relevant change of the motional state.

\section{Conclusion}
Thermally autonomous nonclassical processes are capable of deterministically producing nonclassical quantum states without an external coherent driving force, exploiting only the short time coherence during the interaction of the studied system and its thermal environment. A prominent example of such process, the broadly studied Jaynes-Cummings interaction, can change a thermal state of a quantum mechanical oscillator into a nonclassical state just by passive energy exchange with a two level system. This allows preparation of nonclassical states without the need for ground state cooling of the oscillator, which is a promising prospect for the recently developed field of solid state quantum optomechanics \cite{SolidOpto}.
We have analyzed this thermally autonomous nonclassical process in detail, quantifying the produced nonclassicality by means of logarithmic negativity potential. We have found that, in case of absorption caused by a two-level system in the ground state, the nonclassicality of the produced state grows with its initial temperature - a rather surprising result, as quantum mixedness caused by the high temperature has traditionally been an enemy of nonclassicality. We have proposed experimentally feasible criteria for detecting the generated nonclassicality and suggested experimental verification of the effect for the vibrational modes of trapped ions. We believe that this opens the path for a direct preparation of nonclassical states even for quantum mechanical oscillators with limited control and ability to be cooled.

\medskip

\textbf{Acknowledgements:}
We acknowledge project GB14-36681G of the Czech Science Foundation. L.L. acknowledges IGA-PrF-2015-005.

\section{Appendix A: Maximal violation of criteria in large energy limit}
In the limit of high thermal energy of the oscillator, the number distribution can be expressed as
\begin{equation}\label{}
    P_n \approx \frac{F_n}{\nbar},
\end{equation}
where $F_n=\cos^2 g t \sqrt{n}+ \sin^2 g t \sqrt{n+1}$. The criteria of nonclassicality can be then rewritten as
\begin{equation}\label{}
    n F_n^2 - (n+1)F_{n+1}F_{n-1} > 0.
\end{equation}
This expression can be maximized when $F_n = 2$ and $F_{n-1} =  F_{n+1} = 0$. These conditions require
\begin{eqnarray}
    gt \sqrt{n-1} &=&  \pi k_{-1} + \frac{\pi}{2}, \nonumber \\
    gt \sqrt{n} &=& \pi k_{0}, \nonumber \\
    gt \sqrt{n+1} &=& \pi k_{1} + \frac{\pi}{2}, \nonumber \\
    gt \sqrt{n+2} &=& \pi k_{2} ,
\end{eqnarray}
where $k_{-1}, k_0, k_1, k_2$ and $n$ are arbitrary integers and $gt$ is arbitrary positive parameter. This set of equations can never be satisfied perfectly, but for large $n$ it is possible to satisfy it approximatively by choosing $gt = \pi \sqrt{n}$. As a consequence, for arbitrarily high value of $\nbar$ we can always set the interaction constant in such a way that the produced state maximizes some nonclassicality criterion.


\begin{thebibliography}{99}
\bibitem{NCmixtures}
This can be shown on few simple examples: superpositions of coherent states, the proverbial Schr\"{o}edinger cat states $|\alpha\rangle \pm |-\alpha\rangle$ are highly nonclassical, while their balanced mixture is not. In a similar vein, displaced single photon states $|1;\alpha\rangle = e^{a^{\dag}\alpha - a \alpha^*}|1\rangle$ are nonclassical for all complex $\alpha$, but their Gaussian mixture $\int |1;\alpha\rangle \langle 1;\alpha| e^{-|\alpha|^2} d^2\alpha$ is not. Finally, even the thermal state itself is nothing more than a mixture of highly nonclassical Fock states.
\end{thebibliography}
\end{document}